\documentclass[aps,prl,twocolumn,superscriptaddress,showpacs,preprintnumbers,amsmath,amssymb]{revtex4-1}

\usepackage{graphicx} 
\usepackage{dcolumn}  
\usepackage{csquotes}
\graphicspath{{ps}}
\newcommand{\overbar}[1]{\mkern 1.5mu\overline{\mkern-1.5mu#1\mkern-1.5mu}\mkern 1.5mu}

\usepackage{amsbsy}

\begin{document}

\preprint{\vbox{ \hbox{   }
                      	\hbox{Belle Preprint 2017-01}
                        	\hbox{KEK Preprint 2016-62} 
}}

\title{ \quad\\[1.0cm] First measurement of $\boldsymbol{T}$-odd moments in $\boldsymbol{D^{0} \rightarrow K_{S}^{0} \pi^{+} \pi^{-} \pi^{0}}$ decays}

\noaffiliation
\affiliation{University of the Basque Country UPV/EHU, 48080 Bilbao}
\affiliation{Beihang University, Beijing 100191}
\affiliation{University of Bonn, 53115 Bonn}
\affiliation{Budker Institute of Nuclear Physics SB RAS, Novosibirsk 630090}
\affiliation{Faculty of Mathematics and Physics, Charles University, 121 16 Prague}
\affiliation{Chonnam National University, Kwangju 660-701}
\affiliation{University of Cincinnati, Cincinnati, Ohio 45221}
\affiliation{Deutsches Elektronen--Synchrotron, 22607 Hamburg}
\affiliation{University of Florida, Gainesville, Florida 32611}
\affiliation{Justus-Liebig-Universit\"at Gie\ss{}en, 35392 Gie\ss{}en}
\affiliation{SOKENDAI (The Graduate University for Advanced Studies), Hayama 240-0193}
\affiliation{Gyeongsang National University, Chinju 660-701}
\affiliation{Hanyang University, Seoul 133-791}
\affiliation{University of Hawaii, Honolulu, Hawaii 96822}
\affiliation{High Energy Accelerator Research Organization (KEK), Tsukuba 305-0801}
\affiliation{J-PARC Branch, KEK Theory Center, High Energy Accelerator Research Organization (KEK), Tsukuba 305-0801}
\affiliation{IKERBASQUE, Basque Foundation for Science, 48013 Bilbao}
\affiliation{Indian Institute of Science Education and Research Mohali, SAS Nagar, 140306}
\affiliation{Indian Institute of Technology Bhubaneswar, Satya Nagar 751007}
\affiliation{Indian Institute of Technology Guwahati, Assam 781039}
\affiliation{Indian Institute of Technology Madras, Chennai 600036}
\affiliation{Indiana University, Bloomington, Indiana 47408}
\affiliation{Institute of High Energy Physics, Chinese Academy of Sciences, Beijing 100049}
\affiliation{Institute of High Energy Physics, Vienna 1050}
\affiliation{Institute for High Energy Physics, Protvino 142281}
\affiliation{Institute of Mathematical Sciences, Chennai 600113}
\affiliation{INFN - Sezione di Torino, 10125 Torino}
\affiliation{Advanced Science Research Center, Japan Atomic Energy Agency, Naka 319-1195}
\affiliation{J. Stefan Institute, 1000 Ljubljana}
\affiliation{Kanagawa University, Yokohama 221-8686}
\affiliation{Institut f\"ur Experimentelle Kernphysik, Karlsruher Institut f\"ur Technologie, 76131 Karlsruhe}
\affiliation{Kennesaw State University, Kennesaw, Georgia 30144}
\affiliation{King Abdulaziz City for Science and Technology, Riyadh 11442}
\affiliation{Department of Physics, Faculty of Science, King Abdulaziz University, Jeddah 21589}
\affiliation{Korea Institute of Science and Technology Information, Daejeon 305-806}
\affiliation{Korea University, Seoul 136-713}
\affiliation{Kyoto University, Kyoto 606-8502}
\affiliation{Kyungpook National University, Daegu 702-701}
\affiliation{\'Ecole Polytechnique F\'ed\'erale de Lausanne (EPFL), Lausanne 1015}
\affiliation{P.N. Lebedev Physical Institute of the Russian Academy of Sciences, Moscow 119991}
\affiliation{Faculty of Mathematics and Physics, University of Ljubljana, 1000 Ljubljana}
\affiliation{Ludwig Maximilians University, 80539 Munich}
\affiliation{University of Malaya, 50603 Kuala Lumpur}
\affiliation{University of Maribor, 2000 Maribor}
\affiliation{Max-Planck-Institut f\"ur Physik, 80805 M\"unchen}
\affiliation{School of Physics, University of Melbourne, Victoria 3010}
\affiliation{University of Miyazaki, Miyazaki 889-2192}
\affiliation{Moscow Physical Engineering Institute, Moscow 115409}
\affiliation{Moscow Institute of Physics and Technology, Moscow Region 141700}
\affiliation{Graduate School of Science, Nagoya University, Nagoya 464-8602}
\affiliation{Nara Women's University, Nara 630-8506}
\affiliation{National Central University, Chung-li 32054}
\affiliation{National United University, Miao Li 36003}
\affiliation{Department of Physics, National Taiwan University, Taipei 10617}
\affiliation{H. Niewodniczanski Institute of Nuclear Physics, Krakow 31-342}
\affiliation{Nippon Dental University, Niigata 951-8580}
\affiliation{Niigata University, Niigata 950-2181}
\affiliation{Novosibirsk State University, Novosibirsk 630090}
\affiliation{Pacific Northwest National Laboratory, Richland, Washington 99352}
\affiliation{University of Pittsburgh, Pittsburgh, Pennsylvania 15260}
\affiliation{Punjab Agricultural University, Ludhiana 141004}
\affiliation{Theoretical Research Division, Nishina Center, RIKEN, Saitama 351-0198}
\affiliation{University of Science and Technology of China, Hefei 230026}
\affiliation{Showa Pharmaceutical University, Tokyo 194-8543}
\affiliation{Soongsil University, Seoul 156-743}
\affiliation{Stefan Meyer Institute for Subatomic Physics, Vienna 1090}
\affiliation{Sungkyunkwan University, Suwon 440-746}
\affiliation{School of Physics, University of Sydney, New South Wales 2006}
\affiliation{Department of Physics, Faculty of Science, University of Tabuk, Tabuk 71451}
\affiliation{Tata Institute of Fundamental Research, Mumbai 400005}
\affiliation{Excellence Cluster Universe, Technische Universit\"at M\"unchen, 85748 Garching}
\affiliation{Department of Physics, Technische Universit\"at M\"unchen, 85748 Garching}
\affiliation{Toho University, Funabashi 274-8510}
\affiliation{Department of Physics, Tohoku University, Sendai 980-8578}
\affiliation{Earthquake Research Institute, University of Tokyo, Tokyo 113-0032}
\affiliation{Department of Physics, University of Tokyo, Tokyo 113-0033}
\affiliation{Tokyo Institute of Technology, Tokyo 152-8550}
\affiliation{Tokyo Metropolitan University, Tokyo 192-0397}
\affiliation{University of Torino, 10124 Torino}
\affiliation{Virginia Polytechnic Institute and State University, Blacksburg, Virginia 24061}
\affiliation{Wayne State University, Detroit, Michigan 48202}
\affiliation{Yamagata University, Yamagata 990-8560}
\affiliation{Yonsei University, Seoul 120-749}
\author{K.~Prasanth}\affiliation{Indian Institute of Technology Madras, Chennai 600036} 
  \author{J.~Libby}\affiliation{Indian Institute of Technology Madras, Chennai 600036} 
  \author{I.~Adachi}\affiliation{High Energy Accelerator Research Organization (KEK), Tsukuba 305-0801}\affiliation{SOKENDAI (The Graduate University for Advanced Studies), Hayama 240-0193} 
  \author{H.~Aihara}\affiliation{Department of Physics, University of Tokyo, Tokyo 113-0033} 
  \author{S.~Al~Said}\affiliation{Department of Physics, Faculty of Science, University of Tabuk, Tabuk 71451}\affiliation{Department of Physics, Faculty of Science, King Abdulaziz University, Jeddah 21589} 
  \author{D.~M.~Asner}\affiliation{Pacific Northwest National Laboratory, Richland, Washington 99352} 
  \author{V.~Aulchenko}\affiliation{Budker Institute of Nuclear Physics SB RAS, Novosibirsk 630090}\affiliation{Novosibirsk State University, Novosibirsk 630090} 
  \author{T.~Aushev}\affiliation{Moscow Institute of Physics and Technology, Moscow Region 141700} 
  \author{R.~Ayad}\affiliation{Department of Physics, Faculty of Science, University of Tabuk, Tabuk 71451} 
  \author{V.~Babu}\affiliation{Tata Institute of Fundamental Research, Mumbai 400005} 
% ...
\collaboration{The Belle Collaboration}

\noaffiliation

\begin{abstract}
We report the first measurement of the $T$-odd moments in the decay $D^{0} \rightarrow K_{S}^{0} \pi^{+} \pi^{-} \pi^{0}$ from a data sample corresponding to an integrated luminosity of $966\,{\rm fb}^{-1}$ collected by the Belle experiment at the KEKB asymmetric-energy $e^+ e^-$ collider. From these moments we determine the $CP$-violation-sensitive asymmetry $a_{CP}^{T\textnormal{-odd}} = \left[-0.28 \pm 1.38 ~(\rm{stat.}) ^{+0.23}_{-0.76} ~(\rm{syst.})\right] \times 10^{-3}$, which is consistent with no $CP$ violation. In addition, we perform $a_{CP}^{T\textnormal{-odd}}$ measurements in different regions of the $D^{0} \rightarrow K_{S}^{0} \pi^{+} \pi^{-} \pi^{0}$ phase space; these are also consistent with no $CP$ violation.
\end{abstract}

\pacs{11.30.Er, 13.25.Ft, 14.40.Lb, 13.66.Jn}

\maketitle

\tighten

{\renewcommand{\thefootnote}{\fnsymbol{footnote}}}
\setcounter{footnote}{0}
Standard Model (SM) $CP$ violation, which is due to the Kobayashi-Maskawa mechanism \cite{KM}, is very small $[\mathcal{O}(10^{-3})]$ in interactions involving decays of charm hadrons. Hence, any enhancement with respect to the SM prediction can indicate new physics effects due to particles or interactions not included in the SM \cite{CHARMREVIEW}. The decay $D^{0} \rightarrow K_{S}^{0} \pi^{+} \pi^{-} \pi^{0}$ has a self-conjugate final state that can be used for a precise test of $CP$ symmetry. Due to its large branching fraction of 5.2$\%$ \cite{PDG}, one can isolate a sample of $\mathcal{O}(10^{6})$ decays that allows a test at a precision of $\mathcal{O}(10^{-3})$. This decay has been studied once before \cite{MARKIII} but with a sample of only 140 events. Here, we report the first measurement of the time-reversal ($T$) asymmetry in $D^{0} \to K_{S}^{0} \pi^{+} \pi^{-} \pi^{0}$ decays, which is sensitive to $CP$ violation via the $CPT$ theorem \cite{CPT}. This is the first  $T$ asymmetry measurement for a $D$ meson decay with two neutral particles in the final state, one of which is a $\pi^0$ meson.

For this measurement, we use the method described in Refs. \cite{TODD1,TODD2,TODD3,TODD4}. This method was
used by the FOCUS \cite{FOCUS}, BaBar \cite{BABARTODD1,BABARTODD2}, and LHCb \cite{LHCBTODD} Collaborations for similar measurements of $T$-violating asymmetries in $D^{0}$, $D^{+}$, and $D_{s}^{+}$ decays. The measurement is performed by constructing the scalar triple product
\begin{equation}
C_{T} =  \mathbf{{p}_{1}}\cdot(\mathbf{{p}_{2}}\times \mathbf{{p}_{3}}),
\end{equation}
where $\mathbf{{p}_{1}}$, $\mathbf{{p}_{2}}$, and $\mathbf{{p}_{3}}$ are the momenta of any three of the $D^{0}$ daughter particles. Similarly, $\overbar{C}_{T}$ is defined as the $CP$-conjugate observable with $\overbar{D}^{0}$ daughter particles. There must be at least four particles in the final state for $\mathbf{{p}_{1}}$ to not be co-planar with $\mathbf{{p}_{2}}$ and $\mathbf{{p}_{3}}$ and allow nonzero $C_{T}$. We define two asymmetry parameters as
\begin{align}
	        A_{T} &= \frac{\Gamma(C_{T} > 0) - \Gamma(C_{T} < 0)}{\Gamma(C_{T} > 0) + \Gamma(C_{T} < 0)}, \\
\overbar{A}_{T} &= \frac{\Gamma(-\overbar{C}_{T} > 0) - \Gamma(-\overbar{C}_{T} < 0)}{\Gamma(-\overbar{C}_{T} > 0) + \Gamma(-\overbar{C}_{T} < 0)},
\end{align}
for $D^{0}$ and $\overbar{D}^{0}$, respectively, with $\Gamma$ being a partial decay rate. These asymmetries can be nonzero due to the final state interaction (FSI) effects \cite{BIGI}. These effects are eliminated by taking the difference between $A_{T}$ and $\overbar{A}_{T}$ as
			\begin{equation}
			a_{CP}^{T\textnormal{-odd}} = \frac{1}{2} (A_{T} - \overbar{A}_{T}),
			\end{equation}
for which a nonzero value would be a clear signature of $T$ violation \cite{CPT}.

In this Letter, we also present measurements of $a_{CP}^{T\textnormal{-odd}}$ in nine regions of the final state phase space. The regions are selected to isolate $CP$ eigenstates such as $K_{S}^{0} \omega$, vector-vector (VV) states such as $K^{*\pm}\rho^{\mp}$, Cabibbo-favored (CF) states such as $K^{*-}\pi^{+}\pi^{0}$ and doubly-Cabibbo-suppressed (DCS) states such as $K^{*+}\pi^{-}\pi^{0}$. 

The Belle detector \cite{BELLE} is located at the interaction region of the KEKB asymmetric-energy $e^{+}e^{-}$ collider \cite{KEKB}.  The analysis is performed with a data sample corresponding to an integrated luminosity of 966 fb$^{-1}$ collected at or near center-of-mass energies corresponding to the $\Upsilon(nS)$ ($n$ = 1, 2, 3, 4, 5) resonances, where 74\% of the sample is taken at the $\Upsilon(4S)$ peak. The sub-detectors relevant to this measurement are: a tracking system comprising a silicon vertex detector (SVD) and a 50-layer central drift chamber (CDC), a particle identification system comprising of a barrel like arrangement of time-of-flight (TOF) scintillation counters and an array of aerogel threshold Cherenkov counters (ACC), and a CsI(Tl) crystal-based electromagnetic calorimeter (ECL). These subdetectors are located inside a 1.5~T superconducting magnet.

Samples of Monte Carlo (MC) simulated data are used to optimize the selection criteria and to understand various types of background. The {\tt EvtGen} \cite{EVTGEN} and {\tt Geant3} \cite{GEANT3} software packages are used to generate the events and simulate the detector response, respectively. We also include initial and final state radiation effects \cite{PHOTOS} in the simulation study. 

We reconstruct the final state in $e^{+}e^{-} \to c\bar{c} \to D^{*+} X$ events \cite{CHARGECONJUGATION} in which $D^{*+} \to D^{0} \pi_{\mathrm{slow}}^{+}, ~D^{0} \to K_{S}^{0} \pi^{+} \pi^{-} \pi^{0}$ and $X$ is a collection of particles produced along with the $D^{*+}$ meson. The $\pi_{\mathrm{slow}}^{+}$ meson is so called because its momentum is low compared to the final state particles originating from the $D^{0}$ decay.  We use the charge of $\pi_{\mathrm{slow}}$ to identify whether the accompanying candidate is a $D^{0}$ or a $\overbar{D}^{0}$ meson. 

We require candidate $\pi^{\pm}$ daughters of the $D^{0}$ and $\pi^{+}_{\mathrm{slow}}$ to have a distance of closest approach along and perpendicular to the $e^{+}$ beam direction of less than 3.0~cm and 0.5~cm, respectively;  this removes tracks not originating from the interaction region.  Furthermore, these track candidates need to be positively identified as pions based on the combined information from the CDC, TOF, and ACC. The pion identification requirement has an efficiency of 88\% \cite{PID} with the probability of misidentification of a kaon as a pion candidate of 8\%. We select $K_{S}^{0}\to \pi^{+}\pi^{-}$ candidates from pairs of oppositely charged tracks, both treated as pions. The two tracks are required to have a $\pi\textendash\pi$ invariant mass within $\pm3\sigma$ of the $K_{S}^{0}$ mass \cite{PDG}, where $\sigma$ is the mass resolution. The decay vertex of the $K_{S}^{0}$ candidates is required to be displaced from the $e^{+}e^{-}$ interaction point by a transverse distance of greater than 0.22~cm for momenta greater than 1.5~GeV/$c$, and greater than 0.08~cm for momenta between 0.5 and 1.5~GeV/$c$ \cite{KSHORT}. We select $\pi^{0}$ meson candidates from pairs of photons reconstructed in the ECL. The photons have different minimum energy criteria of 50 MeV, 100 MeV, or 150 MeV, depending on whether they are reconstructed in the barrel, forward endcap, or backward endcap regions of the ECL, respectively. These criteria suppress the beam-related backgrounds, which are typically asymmetric in polar angle. A $\pi^{0}$ candidate is selected when the invariant mass of the photon pair lies between 115 and 145~MeV/$c^{2}$, which covers an asymmetric interval corresponding to 3$\sigma$ about the nominal mass of the $\pi^{0}$ meson \cite{PDG}. We require that $\pi^{0}$ candidates have momentum greater than 350 MeV/$c$ to reduce combinatorial background from random combinations of particles not originating from $D^{0}\to K_{S}^{0} \pi^{+}\pi^{-}\pi^{0}$ decays. We kinematically constrain the $\pi^{0}$ meson to its known mass \cite{PDG} to improve the momentum resolution. We identify a $D^{0} \to K_{S}^{0} \pi^{+} \pi^{-} \pi^{0}$ candidate if its reconstructed invariant mass $(M_{D^{0}})$ is between 1.80 and 1.95~GeV/$c^{2}$.

We select $\pi_{\mathrm{slow}}^{+}$ candidates from the remaining pion candidates in the event that produce at least one hit in the SVD; this requirement reduces the multiplicity of candidates within an event. We form $D^{*+}$ from the selected $D^{0}$ and $\pi_{\mathrm{slow}}^{+}$ candidates. To eliminate $D^{*}$ mesons from $B$ decays, which have different kinematic and topological properties, we require the $D^{*+}$ momentum in the center-of-mass frame to be greater than 2.5~GeV/$c$. A small contamination of 0.015\% and 0.096\% from $B$ and $B_{s}$ events, respectively, is found from MC simulation studies. We define the variable $\Delta M = M_{D^{*+}} - M_{D^{0}}$, where $M_{D^{*+}}$ is the mass of the $D^{*+}$ candidate; this peaks at 145 MeV/$c^{2}$ \cite{PDG} for correctly reconstructed $D^{*+}$ mesons.  We require $\Delta M$ to be less than 150~MeV/$c^{2}$ to suppress the combinatorial background. We perform kinematically-constrained vertex fits for both the $D^{0}$ vertex (using the $\pi^{+}$, $\pi^{-}$ tracks, 
$\pi^{0}$ vertex, and $K_{S}^{0}$ momentum) and the $D^{*+}$ vertex (using the $D^{0}$ momentum and $\pi_{\mathrm{slow}}^{+}$ track). We remove very poorly reconstructed candidates whose vertex fit quality parameter exceeds 1000. We also apply a kinematically-constrained mass fit for the $D^{0}$ meson candidates  to improve the resolution of the momenta of $D^{0}$ daughters. 

Selection criteria are chosen to maximize the significance $S/\sqrt{S+B}$, where $S~(B)$ is the number of MC signal (background) events in the signal region, defined as 144\textendash 147~MeV/$c^{2}$ for $\Delta M$ and 1.82\textendash 1.90~GeV/$c^{2}$  for $M_{D^{0}}$. Two types of backgrounds are significant: (1) \textquoteleft combinatorial\textquoteright~ and (2) \textquoteleft random $\pi_{\mathrm{slow}}^{+}$.\textquoteright~The latter consists of a correctly reconstructed $D^{0}\to K_{S}^{0} \pi^{+} \pi^{-} \pi^{0}$ decay paired with a $\pi_{\mathrm{slow}}^{+}$ candidate that is not from a common $D^{*+}$ parent. The background contributions in the selected data sample are 55\% and 1\% for combinatorial and random $\pi_{\mathrm{slow}}^{+}$ components, respectively. The signal purity is 79\% in the signal region. The selection efficiency estimated from MC simulation is 4$\%$, and the selected data sample contains 1691029 events. 

  The selection results in an average multiplicity of 1.5 $D^{*}$ candidates per event. In events with two or more candidates, we retain for further analysis the one with the smallest $\chi^{2}$ value of the $D^{*}$ vertex. MC studies indicate that this requirement selects the correct candidate in 74\% of the events with multiple candidates. 

We define $C_{T}$ in the $D^{0}$ rest frame as $\mathbf{{p}_{K_{S}^{0}}}\cdot(\mathbf{{p}_{\boldsymbol{\pi^{+}}}} \times \mathbf{{p}_{\boldsymbol{\pi^{-}}}})$ for $D^{0}$ events and $\overbar{C}_{T}$ for $\overbar{D}^{0}$ as $\mathbf{{p}_{K_{S}^{0}}}\cdot(\mathbf{{p}_{\boldsymbol{\pi^{-}}}} \times \mathbf{{p}_{\boldsymbol{\pi^{+}}}})$; the values of $\left|C_{T}\right|$ and $\left|\overbar{C}_{T}\right|$ with other combinations of final state particles are found to yield identical results. To determine $a_{CP}^{T\textnormal{-odd}}$, we first divide the data sample into four categories using the $C_{T}$ value and $\pi_{\mathrm{slow}}$ charge: $(i)$ $D^{0}$ with $C_{T} > 0$, $(ii)$ $D^{0}$ with $C_{T} < 0$, $(iii)$ $\overbar{D}^{0}$ with $-\overbar{C}_{T} > 0$, and $(iv)$ $\overbar{D}^{0}$ with $-\overbar{C}_{T} < 0$.  We then perform a simultaneous maximum likelihood fit to the two-dimensional distributions of $\Delta M$ and $M_{D^{0}}$ to determine $a_{CP}^{T\textnormal{-odd}}$ and yields. The two yields $[(i)$ and $(iii)]$ and two asymmetry parameters ($A_{T}$ and $a_{CP}^{T\textnormal{-odd}}$) of the signal component are floated in the fit.

We model the signal component of the $M_{D^{0}}$ distribution with a probability density function (PDF) that is the sum of a Crystal Ball (CB) function \cite{CB}, a Landau distribution, and two Gaussian functions, with a common value for the Gaussian means and Landau central value. The combinatorial background component is parametrized with a first-order polynomial. The random $\pi^{+}_{\mathrm{slow}}$ component is modeled by the signal PDF. 

The $\Delta M$ signal component is described by a PDF formed from the sum of a CB function, two Gaussians, and an asymmetric Gaussian function. The combinatorial component is parametrized by a PDF that is the sum of an empirical threshold function and a Gaussian function. The threshold function has the form
\begin{equation}
f(\Delta M) = a (\Delta M - m_{\pi})^{\alpha} \exp[-\beta(\Delta M - m_{\pi})],
\end{equation}
where $a$ is the normalization parameter, $\alpha$ and $\beta$ are shape parameters, and $m_{\pi}$ is the mass of the charged pion \cite{PDG}. We observe a small peaking structure in the signal region of the $\Delta M$ combinatorial background distribution that is due to partially reconstructed $D^{0}$ candidates associated with a genuine $\pi_{\mathrm{slow}}^{+}$, such as a correctly reconstructed $D^{*+}\to D^{0}\pi_{\mathrm{slow}}^{+},~D^{0}\to K^{0}_{S}\pi^{+}\pi^{-}$ event combined with a low momentum $\pi^0$ from the rest of the event. We fix the Gaussian parameters and the fraction of Gaussian contribution of the $\Delta M$ combinatorial background PDF to those obtained from the MC sample. The random $\pi_{\mathrm{slow}}^{+}$ component is modeled with the same threshold function as the combinatorial background. 

We calculate signal yields via a two-dimensional unbinned maximum likelihood fit to the values $\Delta M$ and $M_{D^{0}}$. To perform this fit, we include a small correlation term in the PDFs between the width of $\Delta M$ and the value of $M_{D^{0}}$. We parametrize the width of the dominant signal-component Gaussian of $\Delta M$ as
\begin{equation}
		\sigma(\Delta M) = \left.\sigma(\Delta M)\right|_{m_{D^{0}}} + a_{\sigma}(M_{D^{0}} - m_{D^{0}})^{2},
\end{equation}
where $a_{\sigma}$ is a constant and $m_{D^{0}}$ is the known mass of the $D^{0}$ meson \cite{PDG}.

\begin{figure*}[ht!]
\begin{tabular}{cc} 
\includegraphics[width=\columnwidth]{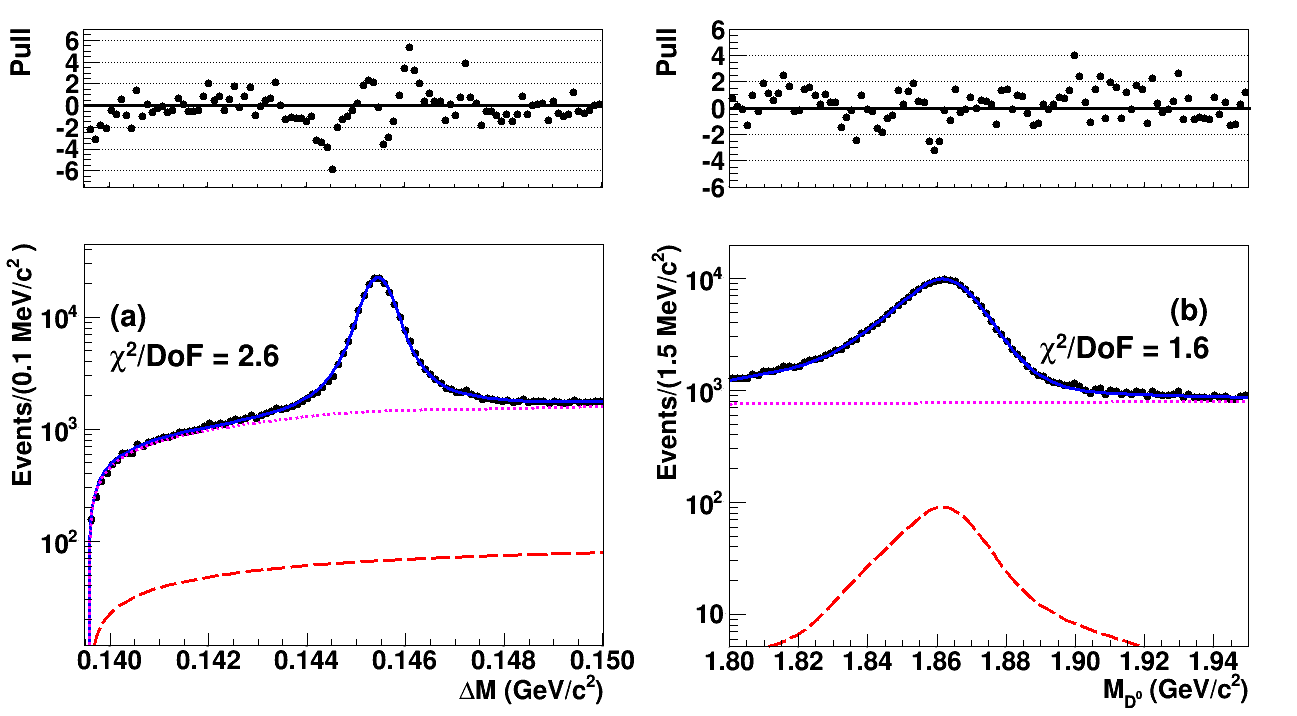}
& \includegraphics[width=\columnwidth]{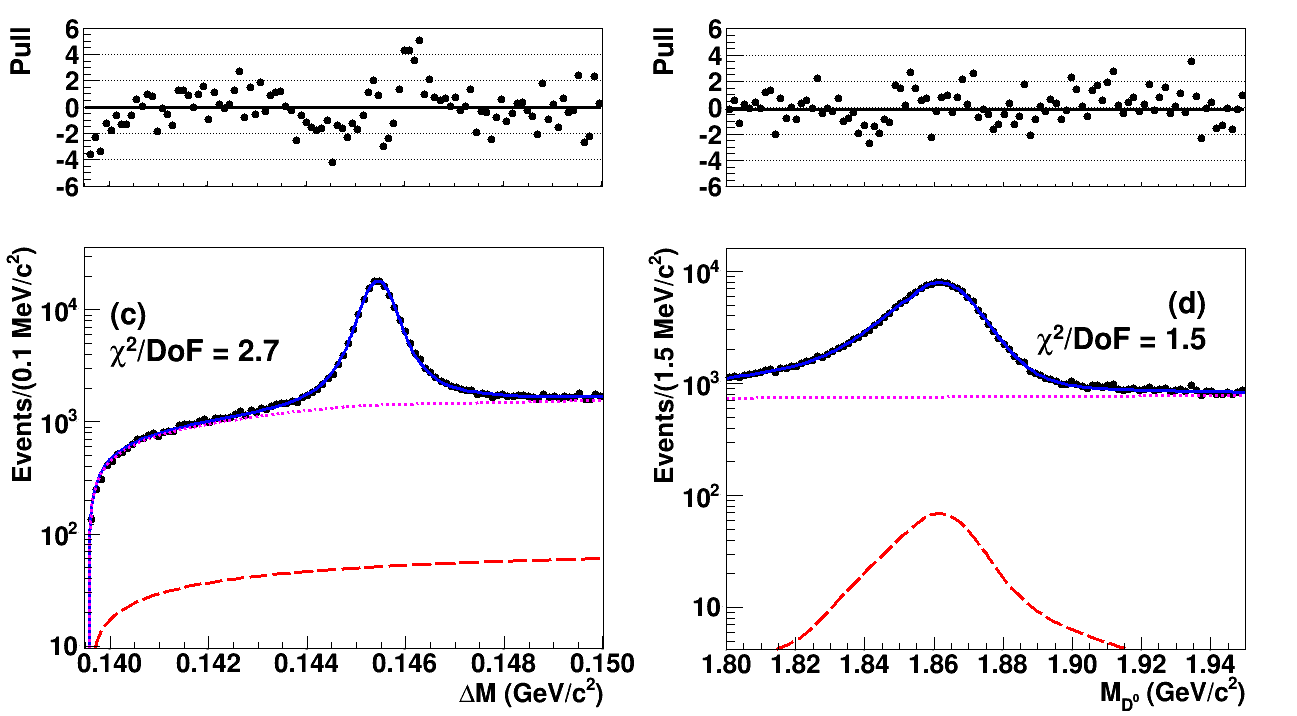}\\
\includegraphics[width=\columnwidth]{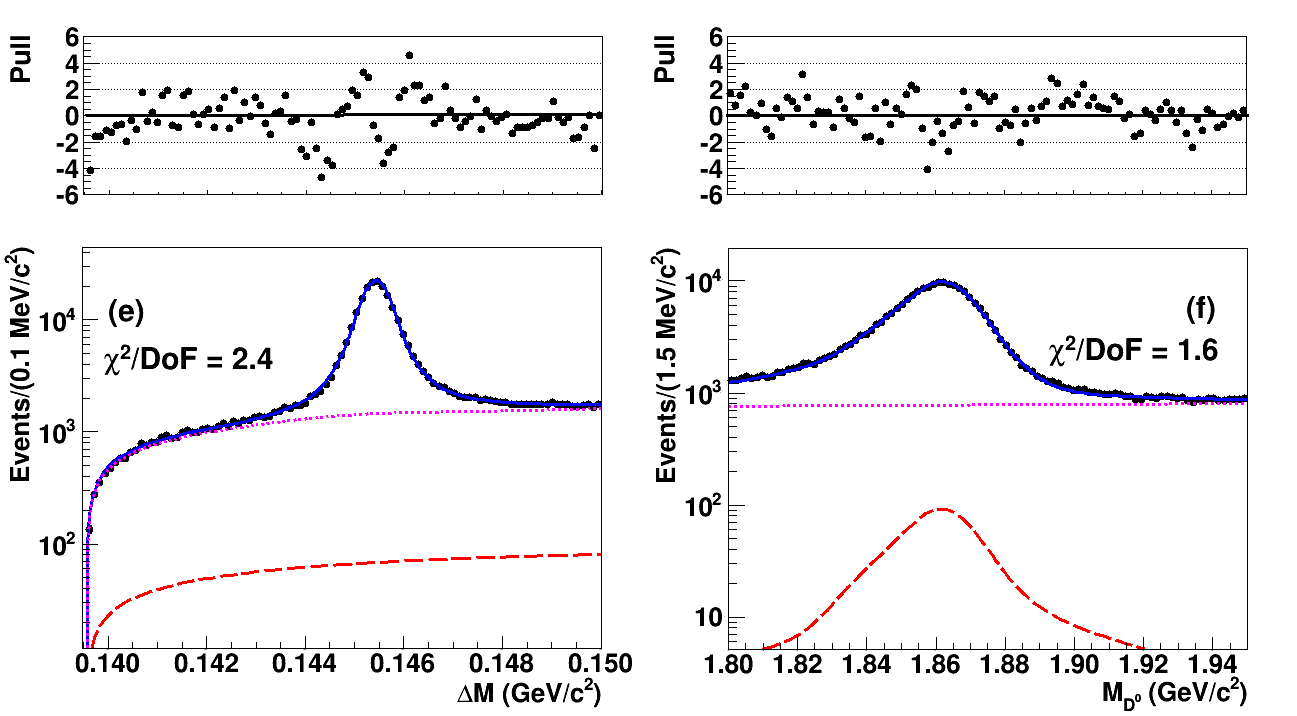}
& \includegraphics[width=\columnwidth]{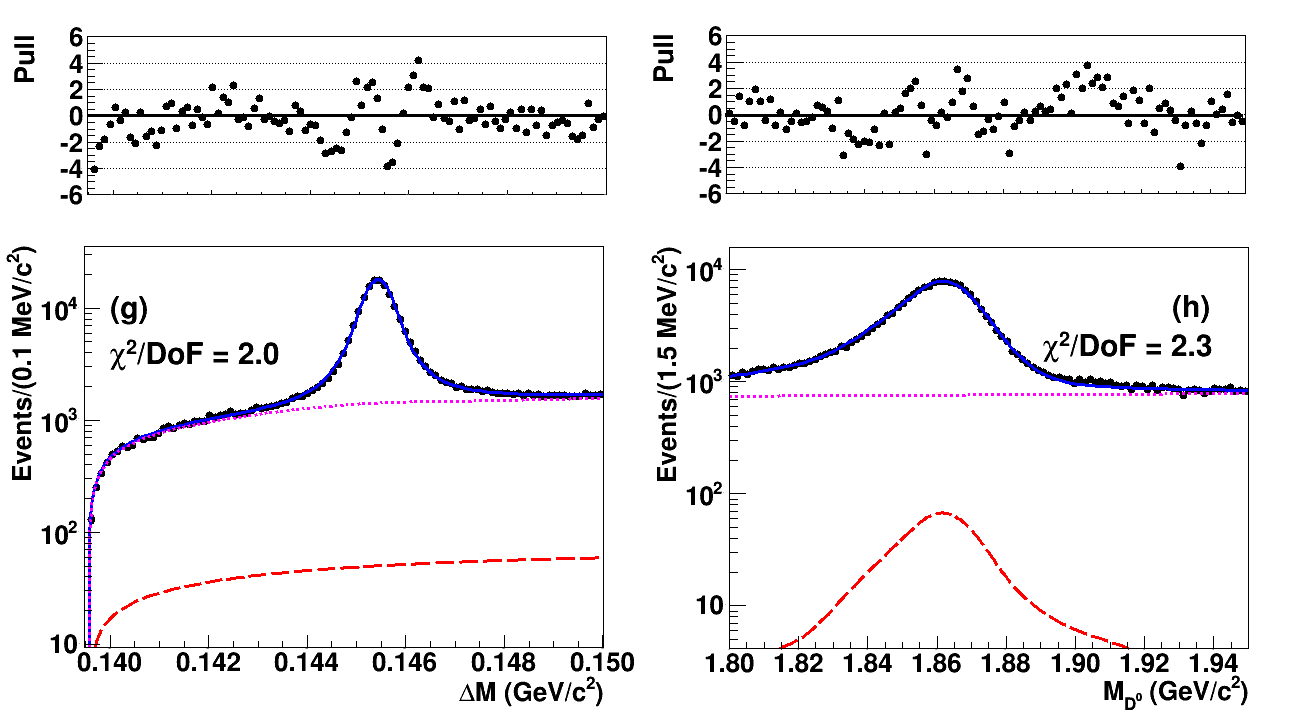}\\ \end{tabular}  \caption{The signal-enhanced logarithmic distributions of (a) $\Delta M$ and (b) $M_{D^{0}}$ for $D^{0}$ with $C_{T} > 0$, (c) $\Delta M$ and (d) $M_{D^{0}}$ for $D^{0}$ with $C_{T} < 0$, (e) $\Delta M$ and (f) $M_{D^{0}}$ for $\overbar{D}^{0}$ with $-\overbar{C}_{T} > 0$ and (g) $\Delta M$ and (h) $M_{D^{0}}$ for $\overbar{D}^{0}$ with $-\overbar{C}_{T} < 0$; the $\Delta M$ distributions have a selection criteria on $M_{D^{0}}$ in the signal region and vice versa. The black points with error bars are the data points and the solid blue curve is the projection of the total signal and background components. The dotted magenta and dashed red curves indicate combinatorial and random $\pi_{\mathrm{slow}}^{+}$ backgrounds, respectively. The normalized residuals (pulls) and the $\chi^{2}/$DoF, where DoF is the number of degrees of freedom, are shown above each plot.} \label{fig:fit} 
\end{figure*}

The background component yields for all four samples are floated independently, but the shape parameters are common for the four categories. In total, there are 21 free and nine fixed parameters in the fit. The parameters fixed from MC are one of the widths of the asymmetric Gaussian, the width and exponent of the CB PDFs in the $\Delta M$ signal component, the normalization parameter $a$ in the threshold PDF, three Gaussian parameters for the peaking structure in the combinatorial background, the relative contribution of the CB and Gaussian functions to the  $M_{D^{0}}$ PDF of the random $\pi_{\mathrm{slow}}^{+}$ component, and the fraction of PDF that contains the correlation in the two-dimensional signal PDF of $\Delta M$ and $M_{D^{0}}$. The signal-enhanced $\Delta M$ and $M_{D^{0}}$ distributions of the data for the four categories are shown in Fig. \ref{fig:fit}, along with the fit projections. The total signal yield obtained from the fit is $744509 \pm 1622$ and the asymmetries are $A_{T} = (11.60 \pm 0.19)\%$ and $a_{CP}^{T\textnormal{-odd}} = (-0.28 \pm 1.38)\times 10^{-3}$, where the uncertainties are statistical. The non-uniform pull for the $\Delta M$ fits is due to the remaining correlation between $\Delta M$ and $M_{D^{0}}$. However, from MC studies we find that this correlation does not cause any bias in the signal yields, in $A_{T}$, nor in $a_{CP}^{T\textnormal{-odd}}$. The large value for $A_{T}$ is due to the FSI effects \cite{BIGI}. The value of $a_{CP}^{T\textnormal{-odd}}$ is consistent with no $CP$ violation.

We divide the $D^{0}\rightarrow K_{S}^{0}\pi^{+}\pi^{-}\pi^{0}$ phase space into nine exclusive regions according to the intermediate resonance contributions. These are (1) $K_{S}^{0}\omega$ ($CP$ eigenstate), (2) $K_{S}^{0}\eta$ ($CP$ eigenstate), (3) $K^{*-}\rho^{+}$ (VV CF state), (4) $K^{*+}\rho^{-}$ (VV DCS state), (5)  $K^{*-}\pi^{+}\pi^{0}$ (CF state), (6) $K^{*+}\pi^{-}\pi^{0}$ (DCS state), (7) $K^{*0}\pi^{+}\pi^{-}$, (8) $K_{S}^{0} \rho^{+} \pi^{-}$ and (9) everything else. Due to the relatively small size of these samples in comparison with the combined one, we reduce the number of free shape parameters to six while fitting the distributions of $\Delta M$ and $M_{D^{0}}$ in each bin. The remaining parameters are fixed to the values obtained from the fit to the combined data sample. The free parameters are the mean and the width of the $\Delta M$ signal component and the four CB parameters for the $M_{D^{0}}$ signal component. The $A_{T}$ and $a_{CP}^{T\textnormal{-odd}}$ values in each bin are listed in Table~\ref{tab:bins}. The results for $a_{CP}^{T\textnormal{-odd}}$ are all consistent with no $CP$ violation. The values of $A_T$ vary significantly due to the different resonance contributions. A value $A_{T} \approx 0$ indicates the presence of a single partial wave, as in bin 2 where the $S$-wave dominates. Values of $A_{T} > 0$ indicate a significant interference between even and odd partial waves as in bins 3 to 9 \cite{RSINHA}. 

\begin{table*}[htb]
\caption{$A_{T}$ and $a_{CP}^{T\textnormal{-odd}}$ values from different regions of $D^{0}\to K_{S}^{0}\pi^{+}\pi^{-}\pi^{0}$ phase space. $M_{ij[k]}$ indicates the invariant mass of mesons $i$ and $j$ [and $k$].}
\label{tab:bins}
\small
\centering
\begin{tabular}
 {@{\hspace{0.1cm}}l@{\hspace{0.2cm}}@{\hspace{0.2cm}}c  @{\hspace{0.2cm}}c@{\hspace{0.2cm}}	@{\hspace{0.2cm}}c@{\hspace{0.2cm}}@{\hspace{0.2cm}}c@{\hspace{0.2cm}}c@{\hspace{0.2cm}}}
\hline \hline
Bin & Resonance	& Invariant mass & $A_{T} (\times 10^{-2})$ & $a_{CP}^{T\textnormal{-odd}} (\times 10^{-3})$\\
    &           & requirement (GeV/$c^{2}$)    &    &       \\
\hline
1	& $\hspace{-0.25 in}K^{0}_S\omega$	&$~~0.762<M_{\pi^{+}\pi^{-}\pi^{0}}<0.802$ & 3.6 $\pm$ 0.5 $\pm$ 0.5 & $-1.7 \pm 3.2 \pm 0.7$	\\
2	& $\hspace{-0.25 in}K^{0}_S\eta$		&$\hspace{0.48 in}M_{\pi^{+}\pi^{-}\pi^{0}} < 0.590$ & 0.2 $\pm$ 1.3 $\pm$ 0.4 & $~~4.6 \pm 9.5  \pm 0.2$	\\
3	& $\hspace{-0.115 in}K^{*-}\rho^{+}$	& $0.790 <M_{K_{S}^{0}\pi^{-}}  <0.994$ & 6.9 $\pm$ 0.3 	$^{+0.6}_{-0.5}$ &$~~0.0 \pm 2.0  ^{+1.6}_{-1.4}$\\
	&					& $0.610 <  M_{\pi^{+}\pi^{0}}~<0.960$	&		&	\\
4	& $\hspace{-0.115 in}K^{*+}\rho^{-}$	& $0.790 < M_{K_{S}^{0}\pi^{+}} < 0.994$ & 22.0 $\pm$ 0.6 $\pm$ 0.6   &$~~1.2 \pm 4.4  ^{+0.3}_{-0.4}$\\
	&					& $0.610 <  M_{\pi^{-}\pi^{0}}~<0.960$	&	& \\
5	& $K^{*-}\pi^{+}\pi^{0}$		& $0.790 < M_{K_{S}^{0}\pi^{-}} < 0.994$ &  25.5 $\pm$ 0.7 $\pm$ 0.5 & $-7.1 \pm 5.2  ^{+1.2}_{-1.3}$\\
6	& $K^{*+}\pi^{-}\pi^{0}$		& $0.790 < M_{K_{S}^{0}\pi^{+}} < 0.994$ & 24.5 $\pm$ 1.0  $^{+0.7}_{-0.6}$  & $-3.9 \pm 7.3  ^{+2.4}_{-1.2}$\\
7	& $K^{*0}\pi^{+}\pi^{-}$		& $0.790 <M_{K_{S}^{0}\pi^{0}}  <0.994$ &  19.7 $\pm$ 0.8 $^{+0.4}_{-0.5}$  & $~~0.0 \pm 5.6  ^{+1.1}_{-0.9}$\\
8	& $\hspace{-0.05 in}K^{0}_{S}\rho^{+}\pi^{-}$		& $0.610 < M_{\pi^{+}\pi^{0}} < 0.960$ & 13.2 $\pm$ 0.9 $\pm$ 0.4   & $~~7.6 \pm 6.1 ^{+0.2}_{-0.0}$	\\
9	& ~Remainder  	& $-$ & 	20.5 $\pm$ 1.0 $^{+0.5}_{-0.6}$		& $~~1.8 \pm 7.4  ^{+2.1}_{-5.3}$\\

\hline \hline
\end{tabular}
\centering
\end{table*}

The sources of systematic uncertainties are the signal and background models, efficiency dependence on $C_{T}$, $C_{T}$ resolution, and potential fit bias. The dominant contribution comes from modelling the signal and background PDFs.  The fixed parameters in the fit not related to the peaking combinatorial background are varied by $\pm 1$ standard deviation from their nominal value obtained from a simulation sample corresponding to the same integrated luminosity as the data; we assign the change in $a_{CP}^{T\textnormal{-odd}}$ as a systematic uncertainty. Without having a suitable control sample to study the peaking component of the combinatorial background, we change the value of the fraction of Gaussian PDF to twice the value found in the MC sample and then to zero. The resulting changes $+0.02 \times 10^{-3}$ and $-0.42 \times 10^{-3}$, respectively, for $a_{CP}^{T\textnormal{-odd}}$ are assigned as a systematic uncertainty. These uncertainties are combined, accounting for correlations among the parameters, to give a total uncertainty of $^{+0.09}_{-0.73}\times 10^{-3}$.

 To study the dependence of the efficiency on $C_{T}$, we calculate the efficiency in 10 bins of $C_T$ between $-0.05$~(GeV/$c$)$^{3}$ and $0.05$~(GeV/$c$)$^{3}$. We find a relative spread of 10\% in efficiency across the bins that varies quadratically as $c_{2}C_{T}^{2} + c_{1}C_{T} + c_{0}$, where $c_{1} =0$ within its statistical limit. This dependence is due to a reduced reconstruction efficiency for low-momentum $D^{0}$ daughters, which tend to have $C_{T}$ values close to zero. We correct the measured $a_{CP}^{T\textnormal{-odd}}$ value for the efficiency dependence and see negligible change because of the symmetry implied by $c_{1} = 0$. We introduce an artificial asymmetry by changing the value of $c_{1}$ by one standard deviation and perform the efficiency correction again. The change in $a_{CP}^{T\textnormal{-odd}}$ of $0.05 \times 10^{-3}$ is assigned as the systematic uncertainty due to the $C_{T}$ efficiency dependence. The parameter $c_{2}$ is found to be different for $D^{0}$ and $\overbar{D}^{0}$ but still compatible within uncertainties. We take the difference of $0.20 \times 10^{-3}$ in $a_{CP}^{T\textnormal{-odd}}$  when applying different efficiency corrections for $D^{0}$ and $\overbar{D}^{0}$ as a systematic uncertainty. The $C_{T}$ resolution follows a Cauchy distribution with zero mean and a half width at half maximum of 1.325~(MeV/$c$)$^{3}$. We add a corresponding smearing to the $C_{T}$ distribution to determine a systematic change in $a_{CP}^{T\textnormal{-odd}}$ due to any asymmetric cross feed between the positive and negative $C_T$ intervals. The variation in $a_{CP}^{T\textnormal{-odd}}$ due to the migration is $0.02~\times 10^{-3}$, which is taken as a systematic uncertainty from this source. We obtain the fit bias systematic uncertainty, which is a multiplicative one, from a linearity test by giving different input values for $a_{CP}^{T\textnormal{-odd}}$ in sets of simulated pseudo-experiments. We find a possible fit-bias uncertainty of $0.28 \times 10^{-5}$. We add all the individual systematic uncertainties in quadrature to obtain a total $a_{CP}^{T\textnormal{-odd}}$ systematic uncertainty of $^{+0.23}_{-0.76}\times 10^{-3}$. 

 In addition to the systematic studies, we perform other cross checks. There is an asymmetry between the number of $D^{0}$ and $\overbar{D}^{0}$ events reconstructed in the data sample due to the forward-backward asymmetry $(A_{FB})$ generated by interference between the virtual photon and $Z^{0}$ boson \cite{AFB}. This production asymmetry, coupled with the asymmetry of the Belle detector, may induce a different reconstruction efficiency as a function of $C_{T}$ for $D^{0}$ and $\overbar{D}^{0}$. This asymmetry is modeled in the MC samples and is found to introduce no bias to the measured value of $a_{CP}^{T\textnormal{-odd}}$. We also measure $a_{CP}^{T\textnormal{-odd}}$ in bins of $\cos{\theta^{*}}$, where $\theta^{*}$ is the polar angle of the $D^{*+}$ with respect to the $e^{+}$ beam direction defined in the center-of-mass system, and find that the results are consistent with the integrated value. To check for any further systematic effect due to detector reconstruction asymmetry for particles of different charges, we compare the momentum and azimuthal angle distributions for $D^{0}$ and $\overbar{D}^{0}$ daughters in data and MC samples and find no significant difference. Furthermore, we study the dependence of the $C_{T}$ distribution on the $D^{*+}$ momentum selection criterion by varying the latter value by $\pm$100 MeV/$c$. No significant change in the shape of the $C_{T}$ distribution is observed. In addition, we estimate the possible contamination from the decay $D^{0} \to \pi^{+} \pi^{-} \pi^{+} \pi^{-} \pi^{0}$, which is an irreducible background, and find that the contribution is negligible.

In summary, we report the first measurement of the $T$-odd moment asymmetry $a_{CP}^{T\textnormal{-odd}} = (-0.28 \pm 1.38 ^{+0.23}_{-0.76}) \times 10^{-3}$  for $D^{0}\to K^{0}_{S}\pi^{+}\pi^{-}\pi^{0}$, consistent with no $CP$ violation. The results in various bins of $K_{S}^{0}\pi^{+}\pi^{-}\pi^{0}$ phase space also show no evidence for $CP$ violation. This result constitutes one of the most precise tests of $CP$ violation in the $D$ meson system \cite{PDG}. The measurement uncertainties are statistically dominated and thus can be improved further with the data from the upcoming Belle II experiment \cite{BELLEII}.


We thank the KEKB group for excellent operation of the accelerator; the KEK cryogenics group for efficient solenoid operations; and the KEK computer group, the NII, and  PNNL/EMSL for valuable computing and SINET5 network support. We acknowledge support from MEXT, JSPS and Nagoya's TLPRC (Japan); ARC (Australia); FWF (Austria); NSFC and CCEPP (China);  MSMT (Czechia); CZF, DFG, EXC153, and VS (Germany); DST (India); INFN (Italy); MOE, MSIP, NRF, BK21Plus, WCU and RSRI  (Korea); MNiSW and NCN (Poland); MES and RFAAE (Russia); ARRS (Slovenia); IKERBASQUE and UPV/EHU (Spain); SNSF (Switzerland); MOE and MOST (Taiwan); and DOE and NSF (USA).


\begin{thebibliography}{99}
\bibitem{CHARGECONJUGATION} Here, and elsewhere in this Letter, charge conjugation is implied.
\end{thebibliography}
\end{document}